%
\def\cf{{\it cf.}}
\def\eg{{\it e.g.}}
\def\ie{{\it i.e.}}
\def\etal{{\it et al.}}
\def\go{\mathrel{\raise.3ex\hbox{$>$}\mkern-14mu\lower0.6ex\hbox{$\sim$}}} 
\def\lo{\mathrel{\raise.3ex\hbox{$<$}\mkern-14mu\lower0.6ex\hbox{$\sim$}}} 
\def\Msun{\ifmmode M_\odot \else $M_\odot$\fi} 
\def\mdot{\ifmmode \dot m \else $\dot m$\fi} 

\ifx\mnmacrosloaded\undefined \input mn\fi



\pubyear{1999}
 \authorcomment{Mon. Not. R. Astron. Soc., in press}  

\begintopmatter  

\title{Cosmic rays from remnants of quasars?}

\author{Elihu Boldt$^1$ and Pranab Ghosh$^{2,3}$}

\affiliation{Laboratory for High Energy Astrophysics, NASA Goddard Space 
Flight Center, Greenbelt, MD 20771, USA}
\vskip 4pt
\affiliation{$^1$Mail Stop: Code 660, e-mail: boldt@lheavx.gsfc.nasa.gov} 
\vskip 4pt
\affiliation{$^2$Mail Stop: Code 662, e-mail: pranab@rufus.gsfc.nasa.gov}
\vskip 4pt 
\affiliation{$^3$Senior NRC Resident Research Associate, on leave from Tata 
Institute of Fundamental Research, Bombay, India,}

\shortauthor{Boldt \& Ghosh}
\shorttitle{Highest Energy Cosmic Rays}

\abstract {Considerations of the collision losses for protons traversing the 2.7 K black body 
microwave radiation field have led to the conclusion that the highest energy cosmic rays, those 
observed at $\geq 10^{20}$ eV, must come from sources within the present epoch.  In light of this 
constraint, it is here suggested that these particles may be accelerated near the event horizons of 
spinning supermassive black holes associated with presently inactive quasar remnants.  The 
required emf is generated by the black hole induced rotation of externally supplied magnetic 
field lines threading the  horizon.  Producing the observed flux of the highest energy cosmic 
rays would constitute a negligible drain on the black hole dynamo.  Observations with 
upcoming air shower arrays and space missions may lead to the identification of candidate 
dormant galaxies which harbor such black holes.  Although the highest energy events observed 
so far are accounted for within the context of this scenario, a spectral upper bound at $\sim
10^{21}$ eV is expected since the acceleration to higher energies appears to be precluded, on 
general grounds.}

\keywords {acceleration of particles - black hole physics - cosmic rays - galaxies: nuclei.}

\maketitle  

\section{Introduction}

The highest energy cosmic ray induced atmospheric air shower yet observed (Bird \etal~1995) 
corresponds to initiation by a $3.2\times 10^{20}$ eV particle.  Yet, due to  
energy degrading ``GZK'' interactions with the pervasive microwave background (Greisen 1966;  
Zatsepen \& Kuz'min 1966), such a baryonic particle (\eg, proton) that energetic could not 
have originated from a distance beyond 50 Mpc; for sources beyond 100 Mpc all proton 
energies are less than $10^{20}$ eV upon arrival here, independent of their initial energy.  
As pointed out by Elbert \& Sommers (1995), accelerating such particles by powerful radio 
sources, the presumed best candidates for achieving these energies, is not possible for this 
particular event since there are no appropriate radio objects in the correct general direction 
that are within an acceptable distance.  What other population, one adequately represented 
locally, should be considered as an alternate candidate?  Although quasars are not a  local 
phenomenon, the {\it present \/} epoch is in fact the 
{\it preferred \/} one for {\it dead  \/} quasars (Schmidt 1978).  
The prospect of such apparently inactive quasar remnants serving as sources of the 
highest energy cosmic rays is explored within the context of their putative underlying 
supermassive black holes.  If the generator driving the high energy particle accelerator is a 
spinning supermassive black hole, the system involved must in all other respects look 
relatively dormant (\eg, no obvious jets).  If not a dead quasar then what  other 
astronomical object could possibly satisfy this requirement?  Considering the apparently 
plausible (albeit speculative) conditions here identified, this ``bottom-up'' scenario appears 
to provide a unique basis for a potential explanation that is remarkably well matched to the 
puzzle.  The more speculative highly interesting ``top-down'' possibility disregarded in this 
present discussion would have the primary particles produced at ultra-high energies in the 
first instance, typically by quantum mechanical decay of some supermassive elementary 
particles related to grand unified theories (Sigl \etal~1995; Kuz'min \& Tkachev 1998). 
 
\section{High energy accelerators}

In the present epoch, there is a drastic paucity of quasars such as the extremely 
luminous ones ($L\go 10^{47}$ ergs/s) evident at large redshifts, those with putative black 
hole nuclei having masses $\go 10^9\Msun $.  Nevertheless, the expected local number of 
{\it dead \/} quasars associated with the same parent population 
(Schmidt 1978; Small \& Blandford 1992; Richstone \etal~1998) 
is expected to be relatively large.  Studies of the X-ray sky indicate 
a pronounced extragalactic cosmic background that arises mainly from accretion powered 
AGN emission at previous epochs (Boldt 1987; Fabian \& Barcons 1992).  The present-epoch 
mass density that has thereby been built up in the form of supermassive black holes 
must now be substantially more than represented by currently active galactic nuclei (Boldt 
\& Leiter 1995).  The black hole mass spectrum for the nuclei of active Seyfert galaxies 
only extends up to $\sim 5\times 10^7 h^{-2}\Msun$ (Padovani, Burg \& Edelson 1990), where
$h\equiv H_0/(100$ km s$^{-1}$ Mpc$^{-1})$. Considering 
a radiative efficiency of $10\%$ for the accretion powered bolometric luminosity of AGNs, 
Chokshi \& Turner (1992) have calculated the mass built up over all earlier epochs and 
thereby estimated that the expected local mass density of  compact galactic nuclei is two 
orders of magnitude greater than that accounted for by Seyfert galaxies.  They conclude that 
more than half this local density is contributed by inactive quasar remnants which are now 
black holes of mass $>10^{8} h^{-2}\Msun$; over $10\%$ of this density is associated with black 
holes of mass $>6\times 10^{8} h^{-2}\Msun$. As emphasized by Chokshi \& Turner (1992), the local 
universe is expected to be well populated by currently inactive remnants of quasars. Based 
on the mass function described by them we have estimated that, for $h\approx 0.5$, the minimum 
number of black holes within 50 Mpc having a mass $>4\times 10^8 \Msun$ is $\sim 40$ and that 
there should be  about a dozen or more quasar remnant black holes of mass $>10^9\Msun$ in this 
volume. These quasar remnant expectations are consistent with being lower limits to the 
number of corresponding supermassive black holes inferred from a recent comprehensive 
study of massive dark objects (MDOs) at the centers of 32 nearby galaxy bulges 
(Magorrian \etal~1998).  In this connection we note that the number of MDOs within 50 
Mpc identified in their sample as being more massive than $10^9\Msun$ is already 8, comparable 
to the total number of Seyfert 1 AGNs out to that distance.  This is a lower limit to the total 
number of such supermassive objects within this volume since their sample of MDOs at the 
centers of nearby galaxy bulges is incomplete, albeit sufficiently large for the correlations 
sought by them (Magorrian 1998).  Using the luminosity function for field galaxies 
(Efstathiou, Ellis \& Peterson 1988) to estimate the incompleteness of their sample suggests 
that the corrected number of present-epoch supermassive MDOs could well be at least an 
order of magnitude greater than that so far observed.
 
	We use the appellation ``dead'' for the quasar remnants considered here in a more 
complete sense than that which is customary in the conventional description of radio 
galaxies as starved/dead qausars (see, \eg, Rees \etal~1982).  The latter merely implies that 
the nuclear luminosity of the object is small, whereas the objects we are referring to are 
inconspicuous as regards jets as well.  Although local dead quasar remnants are manifestly 
underluminous, their underlying supermassive black holes are likely to be sufficiently 
spun-up (\ie, after their many ``Salpeter'' time units of accretion history; see Thorne 1974; 
Rees 1997) to possibly serve as high-energy accelerators of {\it individual \/}  particles. In 
this scenario (\cf~Blandford \& Znajek 1977) externally produced magnetic field lines threading 
the event horizon of such black holes would, by virtue of the induced rotation, generate an 
effective electromotive force characterized by: $emf\propto cBR$, where $B$ is the magnetic field 
strength and $R$  is the effective range over which the concomitant electric field is applicable.  
Scaling to the magnitude for this impressed $B$ field considered by Macdonald and Thorne 
(1982) and taking $R\approx R_g (\equiv GM/c^2)$, the gravitational radius, the expected 
value for the $emf$ is then here estimated as
$$ emf \approx 4.4\times 10^{20} B_4 M_9 \quad {\rm volts},\eqno\stepeq$$			
where  $B_4\equiv B/(10^4 {\rm Gauss})$ and $M_9 \equiv M/(10^9\Msun)$.  As described by Znajek 
(1978) and Blandford (1979), the massive black hole in their model behaves as a battery with 
an emf of up to $10^{21}$ volts, comparable to the above estimate.  We note that radiative 
losses for particles accelerated in such a dynamo are very much larger for electrons than 
protons. Proton energy losses during the phase of acceleration up to $\sim 10^{17}$ eV are 
due principally to pair production in collisions with ambient photons (Blumenthal 1970). For 
the low luminosity objects considered here [\ie, $L\leq 10^{-4} L({\rm Eddington})$], 
however, the associated radiation length is larger than the gravitational radius ($R_g = 
1.5\times 10^{14}M_9$ cm ) of the black hole, for any plausible emission spectrum from the 
accretion disk.  Hence, radiative cascades (basic to the Blandford-Znajek mechanism), 
such as could attend the process of electron acceleration, would not necessarily constitute 
a comparable limiting factor in the present scenario for the acceleration of the relatively 
few (favorably disposed) {\it protons \/} that need to achieve an energy close to that of 
exploiting the full voltage. The dominant energy loss mechanism for protons during the final
acceleration phase involves photomeson production (Blumenthal 1970; Hill \& Schramm 1985).  
Although the total cross-section for this increases somewhat with energy, its value is 
always $< 200$ microbarns at the highest energies of interest (Caso \etal~1998).  For the low 
luminosity objects considered here, then, the mean free path for such inelastic collisions 
is larger than $R_g$.  Thus proton energy losses are not expected to be significant in this 
scenario.

	This proton accelerator might be an intrinsic feature of an underlying dynamo (non 
jet producing) somewhat different from that of the Blandford-Znajek model or, alternately, 
a subsidiary aspect of that mechanism.  The situation we envisage, therefore, is one in 
which the accelerator is not operational in the mode in which quasi-steady conversion of the  
hole's rotational energy into that of luminous radio jets is possible, but where acceleration 
of individual protons to the full voltage may occur, perhaps sporadically, at those episodes 
(possibly of brief duration) when the emf is {\it not \/} shorted out.

\section{Requirements}

	The air-shower events observed at $\geq 10^{20}$ eV (Cronin 1997) correspond to an 
equivalent omnidirectional cosmic ray flux of about one such energetic particle per 
[(kilometer)$^{2}$ - decade].  Assuming this flux arises from a local random distribution of dead 
quasars associated with black hole masses $\go 10^9\Msun$ implies that their average luminosity is 
$<10^{42}$ ergs/s in these energetic particles.  In contrast to the full  electromagnetic process 
discussed by Blandford and Znajek (1977) for effectively extracting much energy (up to a 
rate $\go 10^{45}$ ergs/s for a $10^9\Msun$ object) the particle accelerator aspect considered 
here for sustaining this cosmic ray output ($\go 10^{20}$ eV/particle) would constitute, 
by itself, a relatively negligible power drain on the rotational kinetic energy reservoir 
provided by the putative underlying supermassive ($\go 10^9\Msun$) canonical Kerr hole.  
Replenishing the particles ejected at high energies ($> 10^{20}$ eV) would only require a 
minimal mass input; a luminosity of $10^{42}$ ergs/s in such particles (if protons) corresponds 
to a rest mass loss $< 10^{-5} \Msun$ in a Hubble time.  Since the energy source in this model 
derives from the black hole itself, no further mass input is needed; the principal input 
requirement for the ambient plasma is to generate the magnetic field.  The ejection of the 
energetic protons considered in this model would constitute a leakage of electric charge at a 
rate amounting to less than $0.01\%$ of the total effective current flow associated with the 
Blandford-Znajek dynamo.

\begintable*{1}
\caption{{\bf Table 1.} Upper limits for cosmic ray generation by
nearby massive $(>10^9\Msun)$ dark objects$^a$.}
\halign{%
\rm#\hfil&\qquad\qquad\qquad\rm#\hfil&\qquad\qquad\qquad\rm\hfil#&\qquad\qquad\qquad
\rm\hfil#&\qquad\qquad\qquad\rm\hfil#&\qquad\qquad\qquad\hfil\rm#\cr
\noalign{\vskip 10pt}
{\bf Galaxy}&$M_9$&$\log x$&$\mdot / \mu$
&$B_4 / \mu^{1/2}$&$(emf)_{20} / \mu^{1/2}$\cr
&&&&&\cr
&&&$\leq$&$\leq$&$\leq$\cr
\noalign{\vskip 10pt}
NGC1399&5.2&-1.785&0.028&0.14&3.2\cr
NGC1600&11.6&-2.046&0.051&0.13&6.5\cr
NGC2300&2.7&-2.161&0.067&0.30&3.6\cr
NGC2832&11.4&-1.935&0.040&0.11&5.7\cr
NGC4168&1.2&-2.356&0.105&0.57&3.0\cr
NGC4278&1.6&-1.959&0.042&0.31&2.2\cr
NGC4291&1.9&-1.807&0.029&0.24&2.0\cr
NGC4472&2.6&-2.509&0.145&0.46&5.2\cr
NGC4486&3.5&-2.377&0.110&0.34&5.3\cr
NGC4649&3.9&-2.143&0.063&0.25&4.2\cr
NGC4874&20.8&-2.000&0.045&0.09&8.3\cr
NGC4889&26.9&-1.678&0.021&0.06&6.5\cr
NGC6166&28.4&-1.767&0.027&0.06&7.4\cr
NGC7768&9.1&-1.991&0.044&0.14&5.3\cr
&&&&&\cr
{\bf average}&{\bf 9.3}&&{\bf 0.06}&{\bf 0.23}&{\bf 4.9}\cr
}
\tabletext{\noindent $^a$Based on Table 2 in Magorrian \etal~(1998).
Here, $M_9\equiv M_{MDO}/(10^9\Msun)$, $x\equiv M_{MDO} / M_{Bulge}$,	
$\mdot\equiv [c^2(dM/dt)] / L_{Edd}$, $\mu\equiv$ galaxy bulge mass loss rate 
(\Msun/year) per $10^{12} \Msun$, $B_4\equiv B /(10^4 {\rm Gauss})$, 
$(emf)_{20}\equiv  (emf)/(10^{20} {\rm volts})$}
\endtable

\section{Discussion}

What are the environmental circumstances of the black hole nuclei in dead quasars, 
and are they conducive to sustaining the magnetic fields needed for this model?  We appear 
to encounter a variety of different  situations that can account for the low luminosity of such 
systems.  Why are they so quiescent?  As emphasized by Rees (1997), their environment 
could be almost free of gas, so that very little gets accreted; in this case the accretion rate is 
inefficient, in that the cooling is so slow (because of the low densities) that only a small 
fraction of the binding energy gets radiated before the gas is swallowed, a regime of 
advection-dominated accretion flow (Narayan 1997).  The unusually low luminosities 
associated with the supermassive black holes at the centers of nearby bright elliptical 
galaxies (Loewenstein \etal~1998; Fabian \& Rees 1995; Fabian \& Canizares 1988) have 
been explained in terms of advection-dominated flow (Mahadevan 1997; Narayan 1997), 
even when there is ample ambient gas.  Rather than attempting to address the parameter 
space for this model's potential applicability to all dead quasars or study possible alternate 
models for the microphysics of low-luminosity AGNs, we have considered in some detail 
the phenomenology of the specific massive dark objects (MDOs) in the large sample  
recently reported by Magorrian \etal~(1998) and have investigated the quite general 
constraints they impose on our proposed scenario.  Of the 32 nearby galaxies with MDOs 
considered by Magorrian \etal~(1998) 14 have compact central masses $M_{MDO} > 10^9\Msun$.  
In order to obtain absolute upper limits on the emf  that could be generated by each of these 
putative supermassive black holes we take that the maximum magnetic field possible near 
the event horizon would correspond to pressure equilibrium between it and the infalling 
matter, whereby
$$(B_4)^2 \leq 3.7 \mdot / M_9,\eqno\stepeq$$ 
where \mdot~is the accretion rate in Eddington units, 
($L_{Edd} = 1.26\times 10^{47} M_9$ ergs/s), defined as
$$\mdot\equiv c^2(dM/dt)/L_{Edd}.\eqno\stepeq$$						
The galaxy bulge mass ($M_{Bulge}$), as described by Magorrian \etal~(1998), constitutes the 
ultimate reservoir fixing the maximum accretion rate possible into the central black hole, 
viz:
$$\mdot \leq 4.6\times 10^{-4}\mu/x,\eqno\stepeq$$					
where $\mu$ is the galaxy bulge mass loss rate (\Msun/year) per $10^{12}\Msun$  and 
$$x \equiv M_{MDO} / M_{Bulge}.\eqno\stepeq$$					
Typically, $\mu\go 1$ (Renzini \& Buzzoni 1986; Mathews 1989).
The maximum possible magnetic field is obtained from equations (2) and (4) as
$$B_4\leq 4.1\times 10^{-2}(\mu/x)^{1/2} M_9^{-1/2}.\eqno\stepeq$$			
The largest possible emf, that corresponding to the maximum magnetic field, is then 
obtained from equations (1) and (6) as
$$(emf)_{20} \leq 0.18(\mu/x)^{1/2} M_9^{1/2},\eqno\stepeq$$			
where $(emf)_{20} \equiv emf/(10^{20} {\rm volts})$.

	The limits on $\mdot$, magnetic field and emf obtained from equations (4), (6) \& (7)
 are listed in Table 1 for the 14 most massive nearby MDOs studied by Magorrian \etal~(1998).  
The average maximum magnetic field is $2.3\times 10^3\mu^{1/2}$ Gauss.  
For $\mu\approx 1$ this is substantially less than the magnetic field strength previously 
considered by Macdonald \& Thorne (1982), but comparable to more recent estimates of the likely 
strengths of magnetic fields threading the horizons of accretion-disk fed black holes 
(Ghosh \& Abramowicz 1977).  The average limiting emf ($4.9\times 10^{20} \mu^{1/2}$ volts) is 
remarkably consistent with what would be required to accelerate 
the highest energy cosmic ray yet observed.  What is the prospect of actually achieving 
something close to these limiting values for several supermassive MDOs?  While a well-defined 
specific accretion model that is clearly compatible with the constraints and requirements is still to 
be identified, we find that the possibility of such can not be excluded at this stage, on quite 
general grounds. We anticipate that some version of the current standard view of 
advection-dominated flow (Mahadevan 1997) could be adequate in this respect for many of the MDOs 
considered here.  Exploring the possibly relevant parameter space for such models is beyond 
the scope of this discussion.  However, the margins associated with satisfying the constraints 
listed in Table 1 could well be suggestive of what sort of modification of the theory would be 
needed should the more standard models prove inadequate.  Based on the 14 MDOs listed in 
Table 1, it appears that the generation of cosmic rays more energetic than $\sim 10^{21}$ eV 
is precluded; to a precision commensurate with the statistics of this sample, such a spectral 
cut-off can then be taken as a prediction of our proposed scenario.
   
\section{Outlook}

The Akeno Giant Air Shower Array (AGASA), currently the world's largest, has during 
1990-1997 accumulated an extensive collection of cosmic ray events above $10^{18.5}$ eV, 461 
of which correspond to more than $10^{19}$ eV (Takeda \etal~1998).  Although only six of these 
exceed the critically important threshold of $10^{20}$ eV, the overall data-base already provides 
evidence, albeit still statistically limited, which suggests that the arrival directions and energies 
are most compatible with a scenario in which sources of ultra-high energy protons trace the 
inhomogeneous distribution of luminous matter in the ``local'' present-epoch universe, well 
within 100 Mpc (Medina-Tanco 1999).  If more data at the highest energies confirm this 
preliminary indication it would strongly support models based on cosmic ray proton acceleration 
by present-epoch supermassive black holes and preclude the need for invoking cosmologically 
remote sources of extremely energetic exotic hadrons whose special properties would allow 
them to traverse large distances ($\go 1000$ Mpc) without their energy falling below $10^{20}$ 
eV (Chung, Farrar \& Kolb 1998; Farrar \& Biermann 1998).  The sample of about 100 
extraordinary events with energy $\geq 2\times 10^{20}$ eV expected with the upcoming 
Pierre Auger array (Cronin 1997) will come from nearby sources and, if protons, 
will point accurately to the directions of origin 
(\ie, owing to the correspondingly large particle Larmor radius in the weak 
intergalactic magnetic field).  Candidate galaxies within the acceptable pixels would then be 
searched for stellar-dynamical evidence for central supermassive black hole nuclei (such as the 
MDOs discussed by  Kormendy \& Richstone 1995, Kormendy \etal~1997 and Magorrian \etal~ 
1998), here taken to be indicative of the dead quasar sources of the highest energy cosmic 
rays. If such a correlation is clearly established, and the lack of correlation with strong radio 
sources persists, it would imply that the existence of a black hole dynamo is not a sufficient 
condition for the presence of pronounced jets.  The $OWL$ (Orbiting Wide-angle Light-collectors) 
space borne NASA mission planned for observing, from above, those air showers induced by 
the highest energy cosmic rays is anticipated to have the sensitivity for accumulating an order of 
magnitude more such events than expected with the Auger array (Streitmatter 1998).

\section*{Acknowledgments}

We thank E. L. Turner for confirming that we have appropriately used the result obtained by A. 
Chokshi and him concerning quasar remnants and for his interest in this extension.  It is a 
pleasure to acknowledge valuable discussions with M. Loewenstein and J. Magorrian about 
galaxies having MDOs.  Much appreciated questions posed by the referee have led to substantial 
improvement of our paper.  One of us (E.B.) is particularly grateful to D. Leiter and R. 
Streitmatter for earlier illuminating discussions and to O. W. Greenberg for recent discussions 
on relevant high energy physics and for his considerable encouragement.

\bye